\documentclass[twocolumn,prl,showpacs,preprintnumbers,amsmath,amssymb]{revtex4}

\usepackage{graphicx}
\usepackage{dcolumn}
\usepackage{bm}

\begin{document}

\title{
Spin excitations under fields in an anisotropic bond-alternating quantum $S$=1 chain: contrast with Haldane spin chains
}

\author{M. Hagiwara$^{1,*}$, L. P. Regnault$^2$, A. Zheludev$^3$, A. Stunault$^4$,\\ N. Metoki$^5$, T. Suzuki$^6$, S. Suga$^6$, K. Kakurai$^5$, Y. Koike$^5$, P. Vorderwisch$^7$, and J. H. Chung$^8$}
\altaffiliation[Present address: ]{KYOKUGEN, Osaka University, 1-3 Machikaneyama, Toyonaka 560-8531, Japan}
\affiliation{
$^1$RIKEN(The Institute of Physical and Chemical Research), Wako, Saitama
351-0198, Japan
\\ $^2$CEA-Grenoble, DRFMC-SPSMS-MDN, 17 rue des Martyrs, 38054 Grenoble Cedex 9, France \\ $^3$Condensed Matter Sciences Division, Oak Ridge National Laboratory, Oak Ridge, Tennessee 37831-6393, USA \\
$^4$Institut Laue Langevin, 6 rue J. Horowitz, 38042 Grenoble Cedex 9, France \\ $^5$JAERI, Advanced Science Research Center, Tokai, Ibaraki 319-1195, Japan \\ $^6$Department of Applied Physics, Osaka University, Suita, Osaka 565-9871, Japan \\
$^7$BENSC, Hahn-Meitner Institut, D-14109 Berlin, Germany \\$^8$NIST Center for Neutron Research, National Institute of Standards and Technology, Gaithersburg, Maryland 20899, USA}

\date{\today}

\begin{abstract}
Inelastic neutron scattering experiments on the $S$=1
quasi-one-dimensional bond-alternating antiferromagnet
Ni(C$_9$D$_{24}$N$_4$)(NO$_2$)ClO$_4$ have been performed under
magnetic fields below and above a critical field $H_{\rm c}$ at
which the energy gap closes. Normal field dependece of Zeeman
splitting of the excited triplet modes below $H_{\rm c}$ has been
observed, but the highest mode is unusually small and smears out
with increasing field. This can be explained by an interaction with
a low-lying two magnon continuum at $q_\|=\pi$ that is present in
dimerized chains but absent in uniform ones. Above $H_c$, we find
only one excited mode, in stark contrast with three massive
excitations previously observed in the structurally similar
Haldane-gap material NDMAP [A. Zheludev et al., Phys. Rev. B
$\bf{68}$, 134438 (2003)].
\end{abstract}


\maketitle

 Recent experimental advances allowed studies of a new exciting phenomenon,
namely field-induced Bose condensation of magnons in gapped
quantum magnets~\cite{nikuni}. Particularly interesting is the
case of antiferromagnetic $S=1$ spin chains. For uniform chains
the ground state is an exotic {\it quantum spin liquid} with only
short-range spin correlations and a gap in the excitation spectrum
\cite{haldane}. Antiferromagnetic $S=1$ chains that are not
uniform, but instead feature alternating strong and weak bonds,
are also gapped spin liquids except at a quantum critical
point~\cite{hagiwaranteap}. However, for sufficiently strong
alternation their so-called dimerized ground state is {\it
qualitatively distinct} from the Haldane
state~\cite{affleckhaldane,singhgelfand,katotanaka,yamamoto,narumintenp}.
The differences are significant  yet subtle, and involve the
breaking of the so-called ``hidden'' symmetry. This symmetry is
related to a non-local topological ``string'' order
parameter~\cite{nijsrommelse, kennedytasaki} that is  {\it in
principle}  not observable experimentally. For example,
thermodynamic properties are expected to be almost identical for
Haldane and dimerized cases. Moreover, field-induced Bose
condensation of magnons occurs in both types of spin chains and is
expected to lead them into {\it the same} (in terms of preserved
symmetries of the wavefunction) magnetized high-field
state~\cite{tonegawa}. Thus, at a first glance, for the Haldane
and the dimerized spin chains one can expect the spin dynamics to
be similar in a wide range of external fields. {\em But is it
indeed true?}

 The recently discovered~\cite{escuer} nickel chain
material Ni(C$_9$H$_{24}$N$_4$)(NO$_2$)ClO$_4$, abbreviated to
NTENP, offers an ideal experimental approach to the above problem.
NTENP features  bond-alternating $S$=1 antiferromagnetic chains in
the gapped dimerized phase, as shown by magnetization and ESR
measurenments~\cite{narumintenp}, and neutron scattering
experiments~\cite{zheludevntenp,regnaultntenp}. It is structurally
similar to the Haldane materials NENP
(Ni(C$_2$H$_8$N$_2$)$_2$(NO$_2$)ClO$_4$) and NDMAP
(Ni(C$_5$H$_{14}$N$_2$)$_2$N$_3$(PF$_6$)) whose field behavior was
previously extensively
investigated~\cite{regnaultnenp,zheludevndmap,zheludevndmap2}.
NTENP also has comparable anisotropy and energy scales, which
allows a direct comparison between these systems. In the present
study we performed neutron scattering experiments on 98\%
deuterated NTENP in a wide range of applied fields. We clearly
observed Zeeman splitting of the excited triplet states below
$H_{\rm c}$, a field-induced softening of the spin gap, and the
resulting emergence of antiferromagnetic long range order (LRO)
above $H_{\rm c}$~\cite{tateiwantenp}, similar to what is seen in
NDMAP. However, certain features of the spin excitation spectrum
in NTENP are substantially different. Notable differences in the
spin correlation function are present already at $H=0$, with even
more striking  discrepancies becoming prominent at $H>H_c$.

%
%

First we summarize the crystal and magnetic properties of NTENP.
This compound crystallizes in the triclinic system (space group
$P$$\bar{1}$)~\cite{escuer} with lattice constants
$a$=10.747(1)$\AA$, $b$=9.413(2)$\AA$, $c$=8.789(2)$\AA$,
$\alpha$=95.52(2)$^{\circ}$, $\beta$=108.98(3)$^{\circ}$ and
$\gamma$=106.83(3)$^{\circ}$~\cite{escuer}.  The Ni$^{2+}$ ions
are bridged by nitrito groups along the $a$ axis having two
different bond distances of 2.142(3) and 2.432(6) $\AA$. These
chains are weakly coupled via intervening perchlorate counter
anions. The inversion centers are situated not on the Ni$^{2+}$
ions but on the nitrito groups, so that no staggered components of
the magnetic moments are expected to be retained, thus resulting
in occurence of the long range order above $H_{\rm c}$ at low
temperatures~\cite{tateiwantenp} like in NDMAP~\cite{honda1,honda2}.
The model spin hamiltionian is written
as:
 \begin{eqnarray}
   {\cal H} & = & \sum_{i} [J{\bf S}_{2i-1} \cdot {\bf S}_{2i}+\delta J{\bf S}_{2i} \cdot {\bf S}_{2i+1}-\mu_{\rm B}{\bf S}_i\tilde{g}{\bf H} \nonumber \\ &  & \mbox{}+D(S_i^z)^2],
 \end{eqnarray}
where $J$ is the large exchange constant, ${\bf S}_{2i-1}$, ${\bf
S}_{2i}$ and ${\bf S}_{2i+1}$ the $S$=1 spin operators, $\delta$
the bond-alternating ratio, $\tilde{g}$ the $g$ tensor of
Ni$^{2+}$, $\mu_{\rm B}$ the Bohr magneton and ${\bf H}$ an
external magnetic field, and $D$ is a single ion anisotropy
constant. From the analysis of the magnetic susceptibility data,
the following parameter values were evaluated from a comparison
with numerical calculations~\cite{narumintenp}: the large exchange
constant $J/k_{\rm B}=54.2$~K, the bond alternating ratio
$\delta$=0.45, the single ion anistropy constant $D/J=0.25$
($D/k_{\rm B}$=13.6 K), and $g_\|$=2.14 for the chain direction. A
similar bond-alternating ratio was also evaluated in Inelastic
Neutron Scattering (INS)
measurements~\cite{zheludevntenp,regnaultntenp}. Such a bond
alternating ratio corresponds to the ground state in the dimer
phase. This conclusion was directly confirmed by magnetization,
ESR~\cite{narumintenp} and INS~\cite{zheludevntenp} experiments on
NTENP.

\begin{figure}
\includegraphics[width=3in]{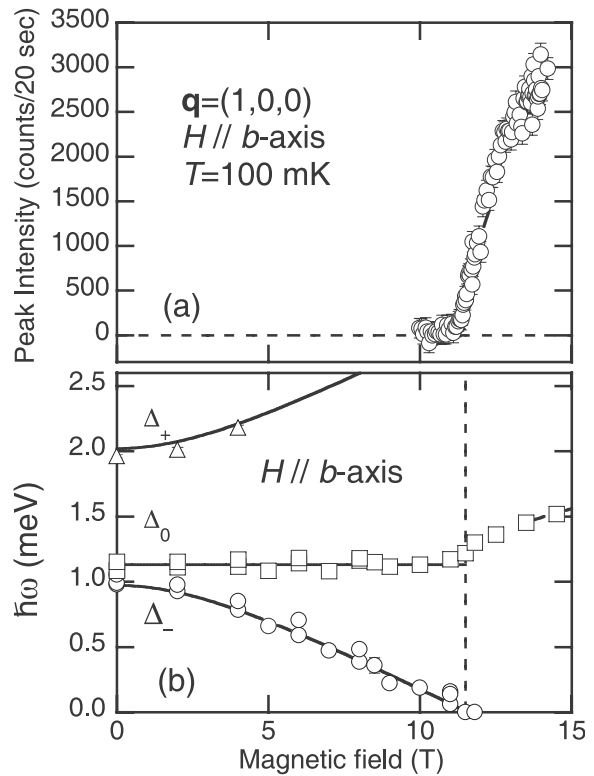}
\caption{(a) Field dependence of the magnetic Bragg peak intensity
at ${\bf q}$=(1,0,0). The solid line is a fit of the experimental
data to the expression $a$($H$-$H_{\rm c}$)$^{2\beta_{\rm c}}$
with $H_{\rm c}$=11.35 T and $\beta_{\rm c}$=0.316 between 11.3 T
and 12 T. The $\beta_{\rm c}$ value is close to the expected 3D-XY
value. The dashed line is a guide for the eyes. (b) Field
dependence of the gap energies measured in NTENP. Solid lines are
calculated triplet branches in a magnetic field and a vertical
thin broken line indicates the critical field $H_{\rm c}$. The
line above $H_{\rm c}$ is a guide for the eyes.}
\label{fig:braggfdep}
\end{figure}

 Three large single crystal samples NTENP with a typical
size of 16$\times$12$\times$4 mm$^3$
were prepared for the current study by the method described in
Ref.~\onlinecite{escuer}. The mosaic spread of the sample was
typically about 1.2$^{\circ}$. The INS measurements were performed
on several three-axis spectrometers installed at different cold and thermal
neutron sources (IN8, IN12, IN14 and IN22 at ILL-Grenoble, V2 at
HMI-Berlin, SPINS at NIST center for neutron research and LTAS at
JAERI). Specific details will be reported elsewhere. The
scattering plane was the $(a^\ast,c^\ast)$ plane and the external
magnetic field was applied to the $b$ axis. Hence, the field
direction was not perpendicular to the chain direction ($a$ axis)
($\gamma=106.83(3)^{\circ}$). In this configuration, the $a^\ast$ was tilted from the $a$ axis by $\approx $30$^{\circ}$ and the field
dependence of the $(1,0,0)$ magnetic Bragg peak measured at 100 mK
clearly indicated a field-induced ordering transition at
$H_c\approx 11.3$~T (Fig.\ref{fig:braggfdep}(a)).

%
\begin{figure}
\includegraphics[width=3in]{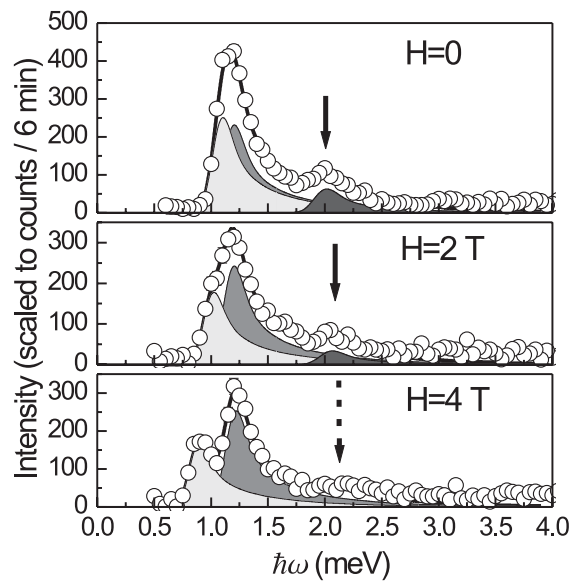}
\caption{Inelastic energy scans near $\bf q$=(1,0,1.5)at 100 mK and $H$=0,
2 and 4 T. } \label{fig:fdepINSbelow}
\end{figure}

\begin{figure}
\includegraphics[width=7.5cm,clip]{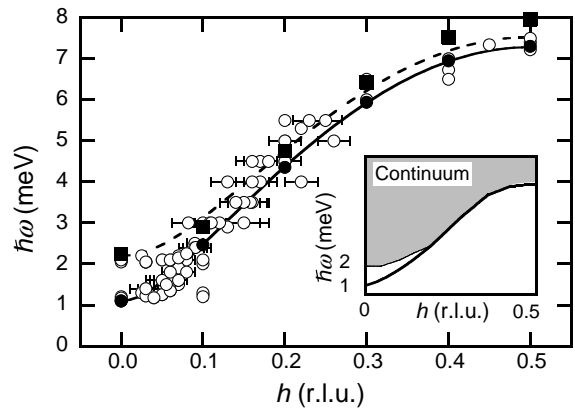}
\caption{ Measured dispersion relation
of magnetic excitations in NTENP (open circles).  Solid symbols show the positions of peaks in the in-plane (squares) and
out-of-plane (circles) dynamic structure obtained by Lanczos
calculations. Solid and broken lines are the results of fitting to the form $(\hbar \omega)^2=\Delta^2+\omega_\mathrm{max}^2 \sin^2{\pi h}$. Inset: excitation continuum (shaded area), as
calculated for NTENP. The lower bound (thin solid line) coincides
with the observed position of the observed $z$-axis mode (the highest excitation mode). The
in-plane modes are shown by a heavy solid line.}
\label{fig:dispersion}
\end{figure}

As a first step in our investigation we characterized the system
at zero field. In agreement with previous studies
\cite{zheludevntenp},  we observed three distinct branches of gap
excitations that appear as resolution-limited peaks in
constant-$q$ scans at the 1D AF zone-center. These data are shown
in symbols in the upper panel of Fig.~\ref{fig:fdepINSbelow}.  Lines represent
contributions of each mode, as calculated in the Single Mode
Approximation (SMA), assuming a parabolic dispersion along the
chain axis and fully taking into account resolution effects, as
was previously done in Ref.~\cite{zheludevntenp}. Each of the
three observed peaks corresponds to a particular member of the
$S=1$ excitation triplet that even at $H=0$ is split by single-ion
anisotropy. The three gap energies were estimated to be
1.06$\pm$0.01, 1.15$\pm$0.01 and 1.96$\pm$0.01 meV, respectively.
The dispersion of the lower modes was measured all the way to the
zone-boundary, as shown by a solid line in the main panel of Fig.\
\ref{fig:dispersion}. It is well described by the form $(\hbar \omega)^2=
\Delta^2+\omega_\mathrm{max}^2 \sin^2{\pi h}$, where the wave
vector transfer $\mathbf{q}=(h,k,l)$, $\Delta=1.1$~meV and
$\omega_\mathrm{max}=7.2$~meV. The higher-energy mode (a broken line) follows a
similar dispersion curve with $\Delta=2.2$~meV and $\omega_\mathrm{max}=7.2$~meV. Apart from a somewhat larger gap energy
and bandwidth, the observed behavior appears to be very similar to
that previously seen in NDMAP\cite{zheludevndmap3}.

 The first key new finding of the present study is
that at $H=0$ the relative intensity of the higher-energy member
of the triplet is anomalously weak. This fact is consistent with
the data shown in Ref.~\onlinecite{zheludevntenp}, but was
previously overlooked. Polarization effects aside, the intensity of
all three components of the triplet in NDMAP scale as $1/\omega$
to a very good approximation. Through application of the 1-st
moment sum rule for $S(\mathbf{q},\omega)$~\cite{hohenberg}, it can be shown
that such behavior directly follows from the SMA, provided that
$(D/J)^2\ll1$. Experimentally, for NTENP (where $D/J$ is actually
smaller than in NDMAP) this scaling is violated, and the relative
intensity of the upper mode is lower than expected by about a
factor of 3. We came to this conclusion after performing
measurements in several Brellouin zones maintaining $q_\|=\pi$, to
eliminate the effect of  polarization-dependent coefficients in
the unpolarized neutron scattering cross section.

The suppression of the upper mode in NTENP becomes even more
apparent in applied magnetic fields. The two lower modes in NTENP
behave almost exactly as in NDMAP for $H$ applied perpendicular to
the spin chains. However, unlike in NDMAP, the upper mode in NTENP
vanishes in relatively modest applied fields of $H\gtrsim$ 4 T.
This is borne out in the constant-$q$ scans shown in the two lower
panels of Fig.\ \ref{fig:fdepINSbelow}. Thus, even well below
$H_c$ the excitation spectra of NTENP and NDMAP are {\it
qualitatively} different. It is interesting that while the
intensities of excitations in NTENP behave anomalously, the field
dependencies of the corresponding gap energies at $H<H_c$ are very
similar to those in NDMAP. These field dependencies measured in
NTENP are plotted in Fig.\ \ref{fig:braggfdep}(b). As in the case
of NDMAP, below $H_c$ these data are well described by a
simple-minded perturbation theory calculation~\cite{golinelli}:
$\Delta_{\pm}(H)=\frac{\Delta_{z}+\Delta_{y}}{2}\pm[(\frac{\Delta_{z}-\Delta_{y}}{2})^2+(g\mu_{\rm
B}H)^2]^{1/2}$ and $\Delta_0(H)=\Delta_{x}$, where $\Delta_{x}$,
$\Delta_{y}$ and $\Delta_{z}$ are the gaps at $H=0$ for
excitations polarized along $x$, $y$ and $z$, respectively. The
best fit is obtained with $\Delta_{z}$=2.0$\pm$0.1 meV,
$\Delta_{x}$=0.97$\pm$0.02 meV, $\Delta_{y}$=1.13$\pm$0.02 and
$g$=2.09.

\begin{figure}
\includegraphics[width=3in]{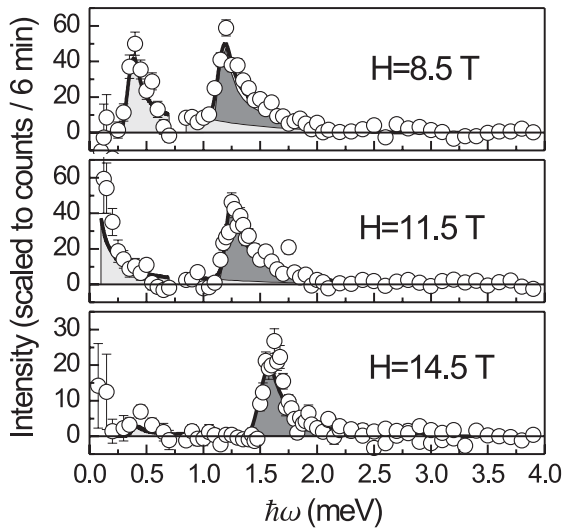}
\caption{Inelastic energy scans at $\bf q$=(1.1,0,1.5) and 100 mK for $H$=8.5 ($<H_{\rm c}$), 11.5 ($\sim H_{\rm c}$) and 14.5 T ($>H_{\rm c}$). }
\label{fig:fdepINSabove}
\end{figure}

The second crucial finding of this work is that at $H>H_c$ there
is {\it only one} low-energy mode at $q_c=\pi$ in NTENP. This can
be seen from Fig.\ \ref{fig:fdepINSabove} that shows constant-$q$
scans collected  below (8.5 T), around (11.5 T) and above(14.5 T)
the critical field ($H_c\sim 11.3$~T). These data are in stark
contrast with similar scans measured in NDMAP (Fig.~1 of
Ref.\onlinecite{zheludevndmap}), where {\it three} distinct modes
are visible at $H>H_c$. As discussed above, the upper mode in
NTENP vanishes well below $H_c$. In addition, in NTENP the
lowest-energy mode that goes soft at $H_c$ {\it does not reappear
at higher fields}, as it does in NDMAP. The data for NTENP at 14.5
T show only a statistically insignificant hint of a peak at 0.4
meV. Scans in several Brilloin zones confirmed that the low-energy
mode is absent at $H>H_c$ in NTENP in all polarization channels.

%
%

Comparing the behaviors of NTENP and NDMAP, we conclude that the
spin dynamics of anisotropic Haldane and dimerized $S$=1 chains
are qualitatively different both below and above the critical
field. This fact is confirmed by a recent Lanczos first-principles
study of the dynamic structure factors in NTENP and NDMAP, based on
actual measured exchange and anisotropy
parameters~\cite{suzukisuga}. The calculated dispersion relations
for NTENP (solid symbols in Fig.\ \ref{fig:dispersion} ) match our
experimental data remarkably well. The solid squares and circles
correspond to peaks in the dynamic structure factors for
excitations polarized parallel to the easy plane ($S^{xx}$ and
$S^{yy}$) or perpendicular to it ($S^{zz}$), respectively. The
calculations also predict that at $H=0$ the intensity of the
out-of-plane $S^{zz}$ mode is only 20\% of that for $S^{xx}$, in
agreement with the neutron results. Finally, the numerical
calculations reproduce the key features at $H>H_c$: three modes
for NDMAP and only one for NTENP.

An understanding of the underlying physical mechanism emerged from
a numerical size-dependence analysis~\cite{takahashi} that can
distinguish between long-lived single-particle excitations and
peaks in multi-particle continua. The key lies in the {\it
violation of translational symmetry} in a dimerized chain. This
symmetry breaking makes the wave vectors $q_\|=\pi$ and $q_\|=0$
equivalent. As a result, in NTENP around $q$=$\pi$ there is a
low-lying continuum that consists of paired $q_\|=0$ and
$q_\|=\pi$ magnons. In the presence of sufficiently strong
anisotropy the upper mode at $q_\|=\pi$ is pushed up in energy and
can enter this continuum. Once this happens, its decay into a pair
of lower-energy excitations becomes allowed by energy-momentum
conservation laws. The upper mode is then no longer a sharp
$\delta$-function of energy, but appears as a asymmetric
finite-width peak on the lower continuum bound (Fig.~\
\ref{fig:dispersion}, inset). The situation becomes progressively
worse when an external magnetic field drives the upper mode to
still higher energies, deeper into the continuum. The two lower
modes remain safely below the lower continuum bound at all times
and are thus not affected. This scenario is reminiscent to that
for the classical antiferromagnet TMMC~\cite{heilmann}. In the
latter, one magnon optic mode interacts with the two magnon mode
to show the avoided crossing. However, the analogy is only
superficial, since, unlike in TMMC, the single-particle
excitations in NTENP are a triplet to begin with.

Nothing of the sort can occur in the uniform Haldane spin chains
of NDMAP, where the magnon gap at $q_\|=0$ is much larger than
that at $q_\|=\pi$ \cite{takahashi}. The lowest-energy continuum
states at $q\|=\pi$ are then three-magnon excitations, and the
corresponding continuum gap is much larger\cite{takahashi}. Even
in the presence of anisotropy all three branches are below the
continuum bounds, and persist as a triplet of long-lived
excitations~\cite{essleraffleck}.

A low-lying continuum is also responsible for the
``disappearance'' of two of the three modes in NTENP at $H>H_c$.
 Numerical calculations show
that, unlike below the critical field, at $H>H_c$ the lower bound
of the continuum in $S^{zz}$ is actually beneath the lowest
single-particle state. Since at any $H>0$ both the highest- and
lowest-energy modes have mixed polarization in the $(y,z)$ plane,
in the high-field phase they {\it both } become subject to decay
into multi-particle states. At $H>H_c$ only the middle mode
survives in NTENP, being polarized along $x$. For this excitation
all decay channels remain closed.


In conclusion, our experimental results clearly demonstrate
fundamental quantum-mechanical differences between the two exotic
spin-liquid phases of uniform and bond-alternating integral spin chains.


\emph{Acknowledgments.---}
This work was in part supported by the Molecular Ensemble research
program from RIKEN and the Grant-in-Aid for Scientific
Research on Priority Areas(B): Field-Induced New Quantum Phenomena
in Magnetic Systems (No.13130203) from the Japanese
Ministry of Education, Culture, Sports, Science and Technology.
Work at ORNL was carried out under DOE Contract No. DE-AC05-00OR22725. Experiments at NIST were supported by the NSF through DMR-0086210 and DMR-9986442.  The high-field magnet at NIST was funded by NSF through DMR-9704257.


\end{document}